\documentclass[conference]{IEEEtran}
\IEEEoverridecommandlockouts
\usepackage{cite}
\usepackage{amsmath,amssymb,amsfonts}
\usepackage{algorithmic}
\usepackage{graphicx}
\usepackage{textcomp}
\usepackage{xcolor}
\usepackage{multirow}
\usepackage{graphicx}
\usepackage{subcaption}
\usepackage{float}
\usepackage{wrapfig}
\def\BibTeX{{\rm B\kern-.05em{\sc i\kern-.025em b}\kern-.08em
    T\kern-.1667em\lower.7ex\hbox{E}\kern-.125emX}}
\begin{document}

\title{A Smartphone based Application for Skin Cancer Classification Using Deep Learning with Clinical Images and Lesion Information\\
}

\author{
\IEEEauthorblockN{Breno Krohling}
\IEEEauthorblockA{Bio-inspired Computing Lab \\
LABCIN - UFES \\
Vitória, Brazil \\
breno.krohling@aluno.ufes.br}
\and
\IEEEauthorblockN{Pedro B. C. Castro}
\IEEEauthorblockA{Bio-inspired Computing Lab \\
LABCIN - UFES \\
Vitória, Brazil \\
pedrobccastro@gmail.com}
\and
\IEEEauthorblockN{Andre G. C. Pacheco}
\IEEEauthorblockA{Bio-inspired Computing Lab \\
LABCIN - UFES \\
PPGI - UFES \\
Vitória, Brazil \\
agcpacheco@inf.ufes.br}
\and
\IEEEauthorblockN{Renato A. Krohling}
\IEEEauthorblockA{Bio-inspired Computing Lab \\
LABCIN - UFES \\
PPGI - UFES \\
Vitória, Brazil \\
rkrohling@inf.ufes.br}
}

\maketitle

\begin{abstract}
Over the last decades, the incidence of skin cancer, melanoma and non-melanoma, has increased at a continuous rate. In particular for melanoma, the deadliest type of skin cancer, early detection is important to increase patient prognosis. Recently, deep neural networks (DNNs) have become viable to deal with skin cancer detection. In this work, we present a smartphone-based application to assist on skin cancer detection. This application is based on a Convolutional Neural Network (CNN) trained on clinical images and patients demographics, both collected from smartphones. Also, as skin cancer datasets are imbalanced, we present an approach, based on the mutation operator of Differential Evolution (DE) algorithm, to balance data. In this sense, beyond provides a flexible tool to assist doctors on skin cancer screening phase, the method obtains promising results with a balanced accuracy of 85\% and a recall of 96\%.

\end{abstract}


\section{Introduction}
The skin cancer occurrence, melanoma and non-melanoma, has increased at a constant rate over the last decades \cite{CancerStats2019}. The World Health Organization (WHO) estimates that 2-3 million non-melanoma cancers and 132,000 melanomas occur every year in the world \cite{WHO2019_HCSK}. The presence of skin cancer is strongly related to the incidence of ultraviolet radiation caused by sunlight exposure \cite{WHO2019_HEUVR}. Due to the lack of pigmentation, caucasian people are under the highest risk \cite{WHO2019_WMRGSK}. Early detection is important to increase patient prognosis \cite{schadendorf2018}.

Several computer-aided diagnoses (CAD) have been proposed to tackle automated skin cancer detection \cite{zhang2020,hekler2019,brinker2019deep,brinker2019comparing,brinker2018,maron2019,esteva2017,bumrungkun2018,alquran2017,Khan2019}. Nowadays, most approaches are based on Convolutional Neural Networks (CNN) trained on dermoscopy images \cite{zhang2020,hekler2019,brinker2019deep,brinker2019comparing,brinker2018,maron2019,esteva2017}. However, in emerging countries such as Brazil, in particular in the countryside \cite{Feng2018}, there is a lack of dermatologists and dermatoscopes\footnote{a medical device that magnifies the lesion for better visualization}, which constraints the use of a CAD system based on dermoscopy images. In this context, smartphones may be useful devices to handle this problem. According to the Ericsson report \cite{ERICSSON2019}, in 2019 the total number of mobile subscriptions around the world was around 8 billion. In Brazil, around 78\% of the population have their own smartphone \cite{IBGE2017}. Therefore, a smartphone-based application to assist clinicians to diagnose skin cancer during the screening process seems to be feasible.

Phillips et al. \cite{Phillips2015} developed an Android application to distinguish melanoma and non-melanoma using a Support Vector Machine (SVM) trained on a dataset composed of 20 images of 3 types of skin lesions. As the model was trained using few samples, its performance is quite limited. Ly et al. \cite{Ly2018} proposed a deep learning model based on convolutional neural network (CNN) for Android and iOS platforms. Their model was tested on the grand challenge PHDB melanoma dataset \cite{ISIC2017} and outperformed the known baseline model in terms of accuracy and computational efficiency. Dai et al. \cite{Dai2019} presented an iOS mobile application for skin cancer also using a CNN. The model was trained on the HAM10000 dataset \cite{Ham10000}, which contains 10,000 dermoscopy images clustered into 7 different types of skin lesions. Both \cite{Ly2018} and \cite{Dai2019} are based on dermoscopy images, which means to use their smartphone application one needs a special dermatoscope attached to it. This is a limitation since this device is expensive and not often available in remote areas. In addition, both applications do not take into account patient demographics.

Pacheco and Krohling \cite{Pacheco2020} proposed a skin lesion classifier for six common skin diseases. The classifier is based on a deep learning model that take into account clinical images and patient demographics, both collected from smartphones. Next, Castro et al. \cite{castroetal2020} developed an  approach based on the mutation operator of Differential Evolution (DE) algorithm to handle the data imbalance problem. In addition, they implement an App for classification of melanoma and non-melanoma skin lesions.
In this paper, we extend their work in the following points:

\begin{itemize}
    \item We include more skin lesions by using the PAD-UFES-20 dataset  \cite{pacheco2020pad}.
    \item We train and validate the model to discriminate between skin cancer and non-skin cancer. The tested CNN model  is used  in a smartphone-application to assist clinicians to diagnose skin cancer during the screening process.
\end{itemize}

\noindent The remainder of this paper is organized as follows: in section 2, we present a literature review of deep learning methods applied to skin cancer classification. In section 3, we describe our previous work and extensions to the data balance approach. In section 4, we present the technologies used to develop the smartphone application. In section 5, we present the experimental results and discussions. Lastly, in section 6, we draw some conclusions.

\section{Related works}
Recently, open datasets containing dermoscopy skin lesion images have significantly increased the number of samples \cite{Ham10000, ISIC2017, ISIC2018}. As a consequence, deep learning models have become viable to tackle skin cancer detection. Remarkable works such as the ones proposed by Esteva et al. \cite{esteva2017} and Brinker et al. \cite{brinker2019comparing} showed that deep learning techniques achieve similar performances to dermatologists. Consequently, many other deep learning approaches have been proposed to classify skin cancer. 

Different works such as Arik et al. \cite{Arik2017}, Shihadeh et al. \cite{Shihadeh2018}, and  Demir et al. \cite{Demir2019} trained and applied common convolutional neural networks (CNNs) architectures, e.g., ResNet \cite{resnet} and Inception-v3 \cite{googleNet}, on dermoscopy images to detect melanoma. Moldovan \cite{Moldovan2019} and Pai and Giridharan \cite{Pai2019} presented similar approachs, however, they classify seven different skin diseases. Other works, such as Alquran et al. \cite{alquran2017} and Mahbod et al. \cite{Mahbod2019} changed the CNN workflow by including a Support Vector Machine (SVM) to work as the model's classifier. In addition, ensemble of CNNs are also common applied for this task \cite{pachecoLearning2020, codella2017, Harangi2018}.

Deep learning approaches to extract the region of interest (ROI) of the lesions were also proposed by De Angelo et al. \cite{de2019skin} and Ali et al. \cite{Ali2019}. Both works are based on the U-Net \cite{ronneberger2015} architecture to segment the skin lesions borders. Nonetheless, they apply the conditional random fields (CRF) \cite{Krhenbhl2011EfficientII} algorithm and a fuzzy filter approach, respectively, as pos-processing methods to improve the segmentation.

Other efforts to improve the performance of the deep learning methods were also proposed. Namozov et al. \cite{Namozov2018} investigated the impact of the activation function on skin lesion classification. Barata et al. \cite{Barata2019} proposed a hierarchical structure, which mimic dermatologists when diagnosing skin lesions, to train and perform a deep learning models. Adegun and Viriri \cite{Adegun2020} proposed a deep convolutional autoencoder based model to melanoma detection. Dos Santo and Ponti \cite{Santos2018} presented an analysis of the impact of dimensionality reduction, colors space contraction, and noisy effects in the feature space in DNNs applied to skin cancer classification. Lastly, Aggarwal et al. \cite{Aggarwal2019}, proposed an attention mechanism that helps CNNs to learn filters that activate pixels of important regions on the skin lesion image to classify three different types of skin cancer.

\section{Methods}\label{sec_methods_and_data}
In this section, we describe an approach to combine clinical images and lesion clinical information using a CNN \cite{Pacheco2020}. Next, we describe the data balancing methods employed in this work.

\subsection{Deep model to classify skin cancer}

Pacheco and Krohling \cite{Pacheco2020} introduced a new dataset composed of clinical images and lesion clinical information as well as an approach to combine both types of data. Each sample in the dataset has a clinical diagnosis, an image, and eight clinical features: the patient's age, the part of the body where the lesion is located, if the lesion itches, bleeds or has bled, hurts, has recently increased, has changed its pattern, and if it has an elevation. We encode the clinical information in 22 variables: 15 bits for region of the body, 1 integer for age, and 6 bits for the remaining features. These features are based on common questions that dermatologists ask patients during an appointment \cite{Pacheco2020}.

In order to combine clinical images and lesion clinical information, Pacheco and Krohling \cite{Pacheco2020} proposed a straightforward mechanism to control the contribution of features extracted from images (FI) and clinical information (CI). We applied the same approach, but now for skin cancer classification. Figure \ref{fig_cnn_arch} shows a schematic diagram of the modified approach proposed in this work:

\begin{figure*}[htbp]
\centering
\includegraphics[width=0.99\textwidth]{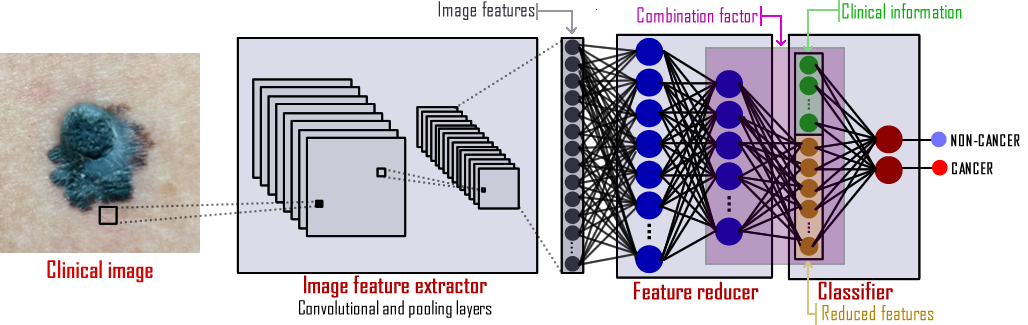}
\caption{The illustration of the model proposed by Pacheco and Krohling \cite{Pacheco2020}. In this work, we modified the last layer for skin cancer classification}
\label{fig_cnn_arch}
\end{figure*}

It is possible to assign more importance for FI or CI by changing the number of features of each one. As the number of clinical data $N_{CI}$ is fixed, one can manipulate the number of features extracted from the image $N_{FI}$. In Equation \eqref{eq_feat_comb} is described how  to calculate the $N_{FI}$ given the $N_{CI}$ and the contribution factor ($\lambda$) of $N_{CI}$ from all the features.

\begin{equation}
    N_{FI} = \dfrac{N_{CI}}{1-\lambda} - N_{CI}, \;\lambda \in [0, 1].\label{eq_feat_comb}
\end{equation}





\subsection{Data Balancing}
A dataset is imbalanced when the number of samples for each class is not uniform distributed among the classes. Classifiers tend to perform worse on imbalanced dataset since they are designed to generalize from data \cite{Akbani2004}. To deal with imbalanced data, we have applied the standard data balancing technique weighted loss function. In addition, we present a competitive approach based on the mutation operator of Differential Evolution (DE). In the following, we described the methods used for data balancing.

\subsubsection{Weighted loss function}

This technique does not change the frequency of samples on datasets. It consists of using a weighted loss function based on a strategy that penalizes miss classification of minority classes. In this paper, we applied the weighted cross-entropy as a loss function. The weight assigned to each label is described by:

\begin{equation}
    W_{i} = \dfrac{N}{n_{i}}\label{eq_feat}
\end{equation} where $N$ is the total of samples and $n_{i}$ is the number of samples of class $i$.

\subsubsection{Differential Evolution (DE)}

Inspired by the mutation operator from the differential evolution algorithm \cite{Storn1997}, the proposed method combines 3 images resulting in a new image. The operator is defined as follows:

\begin{equation}
    X_4 = X_1 + \alpha(X_2 - X_3)
\end{equation} where $X$ is a set of images and ${\alpha}$ is a factor ranging from -0.5 to 0.5, a new value for ${\alpha}$ is chosen in each combination according to a uniform probability distribution. Regarding clinical information used, for each combination generated, the clinical information is randomly chosen between one of the three base images. This technique is applied only for the same kind of skin lesion, which belongs to the same class.

\section{App development}
In order to assist on skin cancer classification, we developed a multi-platform smartphone application. The app's purpose is to assist clinicians who have no or low dermatological experience or do not have access to a dermatoscope. Using the app, clinicians may prioritize patients with possible skin cancer on screening process, leading them to a specialist.

\begin{figure}[htbp]
\centering
\includegraphics[width=0.35\textwidth]{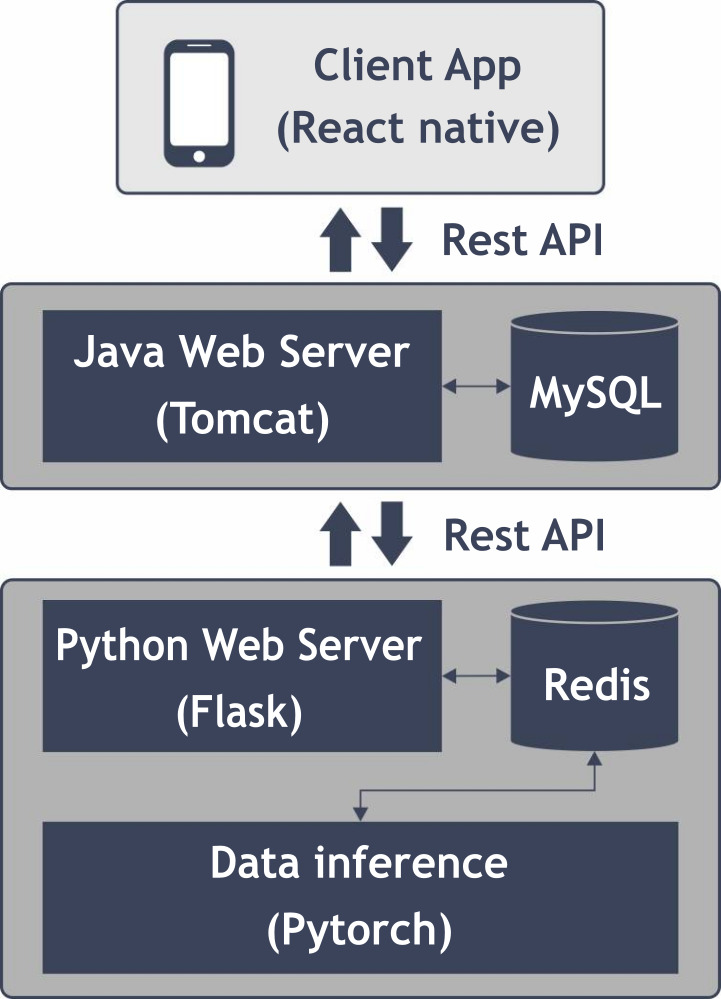}
\caption{Schematic diagram of the smartphone app to skin cancer detection}
\setlength{\belowcaptionskip}{-10pt}
\label{fig_apd_app}
\end{figure}

Embedding a CNN in a smartphone presents two main requirements: 1) the weight's size that can be too large and does not fit on the device's memory; and 2) the need of computational resource to perform the model. Since not all smartphones can fulfill these requirements, we decided to deploy the CNN model on a server. Figure \ref{fig_apd_app} shows a schematic diagram of the developed system.

On the client side, we have a mobile application developed using \textit{React Native}\footnote{https://facebook.github.io/react-native/docs/getting-started} framework, and \textit{Expo SDK}\footnote{https://docs.expo.io/versions/latest/}. The application sends skin lesion images along with their clinical information to the server. The server performs the CNN model and replies the diagnosis prediction. Finally the app displays it on the screen. 

\begin{figure}[H]
\centering
\begin{subfigure}{.20\textwidth}
  \centering
  \includegraphics[width=.8\linewidth]{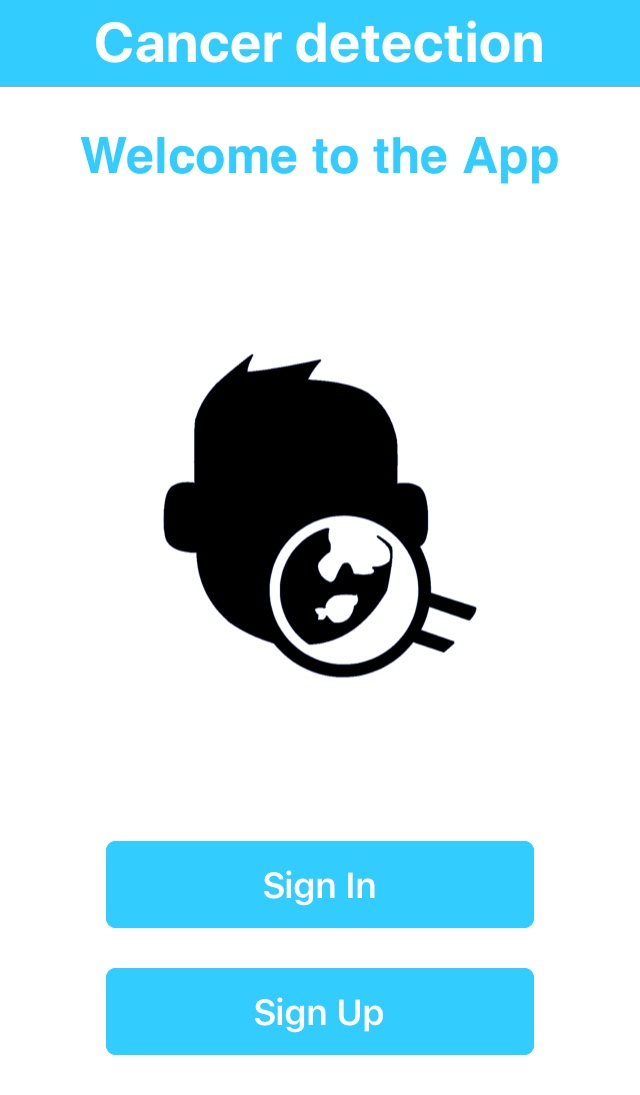}
  \caption{App home screen}
  \label{fig_app_home}
\end{subfigure}%
\begin{subfigure}{.20\textwidth}
  \centering
  \includegraphics[width=.8\linewidth]{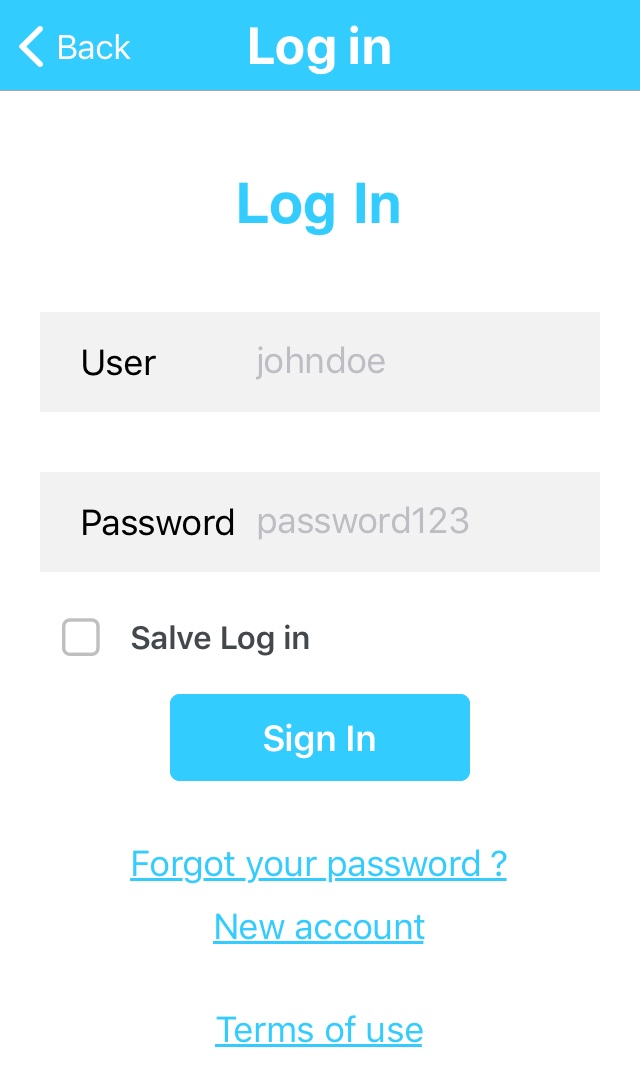}
  \caption{App log in}
  \label{fig_app_login}
\end{subfigure}
\caption{App's home screen and log in screen}
\label{fig_app_1}
\end{figure}
\begin{figure}[H]
\centering
\begin{subfigure}{.20\textwidth}
  \centering
  \includegraphics[width=.8\linewidth]{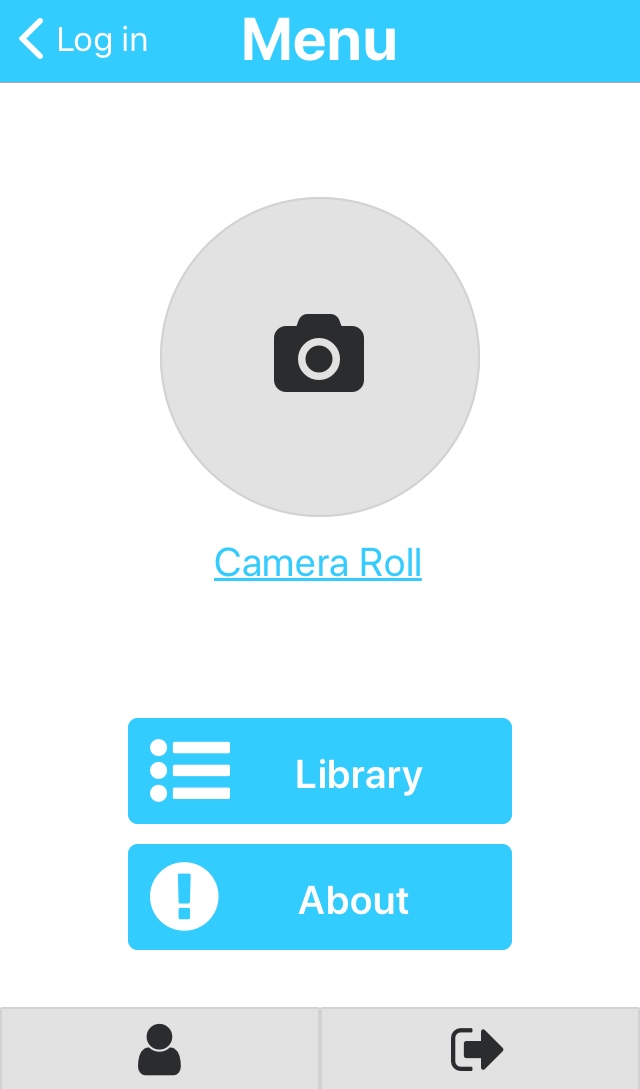}
  \caption{App menu}
  \label{fig_app_menu}
\end{subfigure}%
\begin{subfigure}{.20\textwidth}
  \centering
  \includegraphics[width=.8\linewidth]{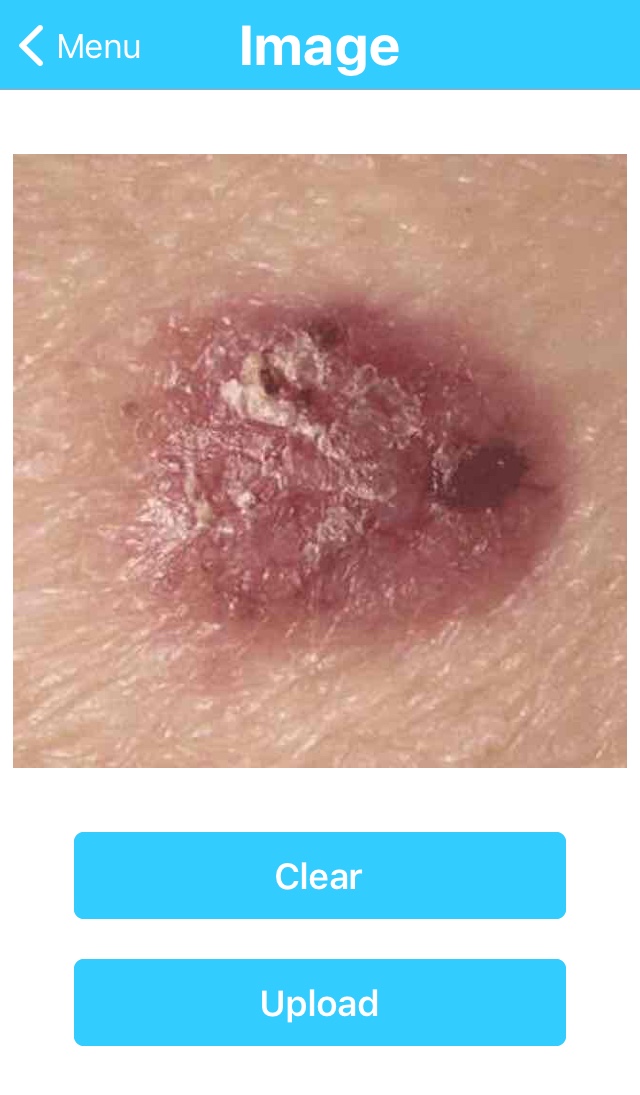}
  \caption{App image acquisition}
  \label{fig_app_pic}
\end{subfigure}
\caption{App's menu screen and image acquisition's screen}
\label{fig_app_2}
\end{figure}
\begin{figure}[H]
\centering
\begin{subfigure}{.20\textwidth}
  \centering
  \includegraphics[width=.8\linewidth]{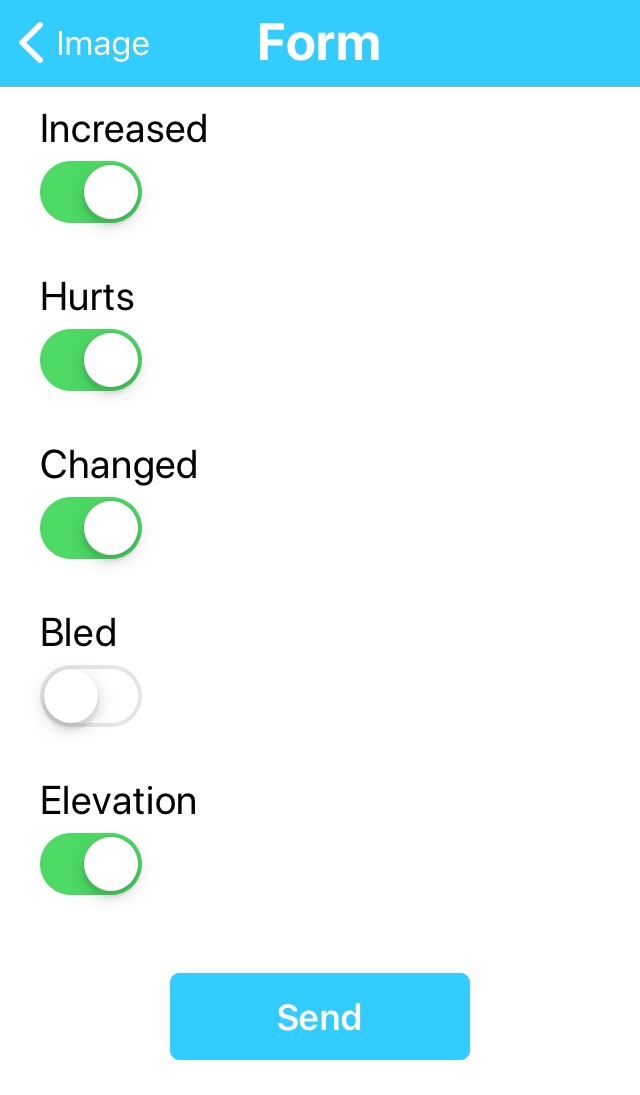}
  \caption{App clinical information}
  \label{fig_app_form}
\end{subfigure}%
\begin{subfigure}{.20\textwidth}
  \centering
  \includegraphics[width=.8\linewidth]{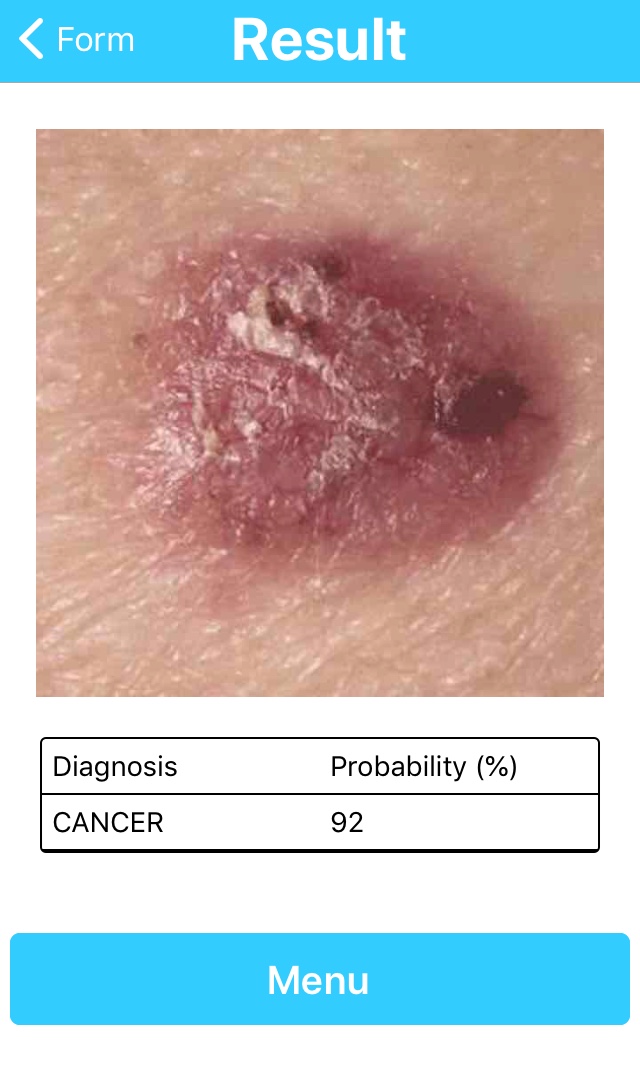}
  \caption{App result}
  \label{fig_app_result}
\end{subfigure}
\caption{App's clinical information screen and result's screen}
\label{fig_app_3}
\end{figure}

The first back-end layer is based on the java web server \textit{Tomcat} that implements a \textit{Rest API} to be consumed by the user as a service. All user information, for log purpose, is stored in a \textit{MySQL} database. The second layer is based on \textit{Flask}\footnote{https://flask-doc.readthedocs.io/en/latest/}, a framework based on \textit{Python} that is designed for micro applications. It makes a direct execution of the machine learning models, which were developed also in \textit{Python}. Every request for processing a new clinical image with its clinical information that arrives at \textit{Flask} is queued in \textit{Redis} \footnote{https://redis.io/}, a \textit{NoSQL} key-based database. If no data is being processed then the first available data of the \textit{Redis} queue is read and sent to a previous trained model. The result of the model is then stored on \textit{Redis} that will be further consulted by the user. Next is presented screenshots of the application to illustrate its workflow. Figure \ref{fig_app_1} shows the main screen and the log in screen, respectively. Next, Figure \ref{fig_app_2} shows the menu and the image upload process, respectively. Last, Figure \ref{fig_app_3} shows the form to collect clinical information and the image of the lesion itself with the diagnosis prediction, respectively.

\section{Results and discussions}
In this section, we present the dataset used to train the CNN, the criteria used in the networks evaluation, a visualization of the features extracted from the images, the the results obtained from simulations and some discussions.

\subsection{Dataset}

The PAD-UFES-20 dataset \cite{pacheco2020pad} used in this work is composed by a set of common skin diseases. Each sample within the dataset is composed by a clinical image and a set of metadata related to the lesion. In our experiments, skin cancer consists of  Melanoma (MEL), Basal Cell Carcinoma (BCC), and Squamous Cell Carcinoma (SCC). The non-cancer skin lesions are  Actinic Keratosis (ACK), Seborrheic Keratosis (SEK), and Nevus (NEV). We added one more disease class labeled as Others, which includes lesions that were not represented in the previous classes. Next, we split the data into skin cancer and non-cancer as presented in Table \ref{tab_dataset_melanoma}.

\begin{table}[ht!]
\begin{center}
\begin{tabular}{cccc}
\hline
{\textbf{Disease*}} & \textbf{Images*} & \textbf{Diseases} & \textbf{Images}\\
\hline
MEL & 67 & \multirow[c]{3}{*}{Cancer} & \multirow[c]{3}{*}{658}\\

BCC & 442 & & \\

SCC & 149 & & \\  \hline

ACK & 543 & \multirow[c]{4}{*}{Non-cancer} & \multirow[c]{4}{*}{1399}\\

SEK & 215 & & \\
 
NEV & 196 & & \\

OTHERS** & 445 & & \\

\hline
\textbf{Total} & \multicolumn{3}{c}{2057} \\
\hline
\multicolumn{4}{l}{*Original Dataset. **New Class}
\end{tabular}
\caption{The frequency of each skin lesion on the extended dataset}
\label{tab_dataset_melanoma}
\end{center}
\end{table}

\subsection{Evaluation criteria}

As evaluation criteria, we aimed first at a high recall, followed by a high accuracy, and last for a high precision. This choice is justified since the recall is directly related to the number of false negative, i.e., the number of skin cancers classified by the network as non-cancer. A false negative is the worst scenario for skin cancer classification since the clinician assumes that the lesion is a non-cancer. The precision is related to the number of false positive, meaning the number of non-cancer lesions classified as skin cancer by the network. In this case, although the patient would be worried, the clinician will send the patient to a specialist.

\subsection{Results}\label{subsec_resu_mel}

The results are divided  according to the type of simulation performed. First, we present a sensitivity study used to find the best setup for the model. Second, we review the impact of clinical information combined with image on the CAD performance. Finally, we investigated the impact of data balancing on the model's performance.

For all tests, a ResNet50 was trained using the architecture described in Pacheco and Krohling \cite{Pacheco2020}, i.e., combining features extracted from the images with lesion clinical information using a 5-fold cross-validation. ResNet50 was used due to its effective performance \cite{Pacheco2020}. We performed the training phase for 100 epochs using Stochastic Gradient Descent (SGD) optimizer with a learning rate equal to 0.01 that decreases by a factor of 1/2 or 1/5 every 20 epochs, alternately. We applied a standard data augmentation \cite{Perez2018} and used the presented techniques to deal with imbalanced dataset. All images were resized to 224×224×3. The evaluation metrics were: balanced accuracy (BACC), precision (PR), recall (REC), and F-measure. The F-measure was used because of the imbalanced dataset.

\subsubsection{Best Setup}\label{subsec_best_setup}

In order to find the best setup, we used standard data augmentation and weighted loss to deal with the imbalanced dataset, as proposed in \cite{Pacheco2020}. The goal is to find the network's best configuration of hyperparamters, so we can apply it to further experiments. In order to increase the system's recall, we also introduced the use of F-measure defined by: 

\begin{equation}
F_\beta = (1 + \beta^2)\dfrac{PR . REC}{(\beta^2 . PR) + REC}\label{form_fmeasure}
\end{equation}
We carried out a sensitivity analysis, changing the value of $\beta$ in \eqref{form_fmeasure} along with 5 combinations of features (C) extracted from the image (FI) and clinical information (CI). Table \ref{tab_sensi_config} presents all 5 combinations of FI and CI.

\begin{table}[htbp]

\begin{center}
\begin{tabular}{ccccc}
\hline
\textbf{C} & \textbf{FI x CI} & \textbf{$N_{FI}$} & \textbf{$N_{CI}$} \\
\hline
1 & 90$\%$ x 10$\%$ & \multirow[c]{5}{*}{22} & 198 \\
2 & 80$\%$ x 20$\%$ &  & 88 \\
3 & 70$\%$ x 30$\%$ &  & 51 \\
4 & 60$\%$ x 40$\%$ &  & 33 \\
5 & 50$\%$ x 50$\%$ & & 22 \\
\hline
\end{tabular}
\caption{Sensitivity analysis taking into account the importance of the FI and CI}
\label{tab_sensi_config}
\end{center}
\end{table}

From these experiments, we obtained the best setup with $\beta$ equal to 7 and 70\% of FI and 30\% of CI.. The metrics using this setup is a BACC of 85.50 $\pm$ 2.47, a precision of 65.09 $\pm$ 5.80 and a recall of 96.42 $\pm$ 2.77.

\subsubsection{The impact of clinical information}

For the study of the impact of clinical information, the number of features extracted from image is equal to the best results obtained in Sec. \ref{subsec_best_setup}. Also, we simulated with the five values of beta used previously in order to find the best results as listed in Table \ref{tab_beta_sem_dados}. From Table \ref{tab_beta_sem_dados} and Sec. \ref{subsec_best_setup}, we can compare the performance of the network with and without clinical information. Table \ref{tab_clin_study} presents these results. From Table \ref{tab_clin_study}, we can notice that the use of clinical information provided an average increase in BACC, precision and recall of 1.41\%, 1.14\% and 2.39\%, respectively.

\begin{table}[htbp]

\begin{center}
\begin{tabular}{cccc}
\hline
\multirow[c]{2}{*}{\textbf{Beta}} & \multicolumn{3}{c}{\textbf{Metrics}} \\

& \textbf{\textit{BACC}} & \textbf{\textit{PR}}& \textbf{\textit{REC}} \\
\hline
{1} & {88.12 $\pm$ 1.39} & {81.64 $\pm$ 4.06} & {85.67 $\pm$ 4.07} \\

{3} & {88.23 $\pm$ 3.01} & {75.80 $\pm$ 6.75} & {90.74 $\pm$ 2.57} \\

{5} & {84.09 $\pm$ 2.73} & {63.95 $\pm$ 5.23} & {94.03 $\pm$ 3.13} \\

{7} & {85.83 $\pm$ 1.62} & {68.53 $\pm$ 2.60} & {91.94 $\pm$ 2.42} \\

{10} & {85.83 $\pm$ 3.26} & {68.5 $\pm$ 3.56} & {91.94 $\pm$ 5.05} \\
\hline
\end{tabular}
\caption{Model's performance for F-measure regarding varying beta without clinical data in the classification process}
\label{tab_beta_sem_dados}
\end{center}
\end{table}

\begin{table}[htbp]
\centering

\begin{tabular}{cccc}
\hline
\multirow[c]{2}{*}{\textbf{Beta}} & \multicolumn{3}{c}{\textbf{Metrics}} \\

& \textbf{\textit{BACC}} & \textbf{\textit{PR}}& \textbf{\textit{REC}} \\
\hline
{3*} & {85.50 $\pm$ 2.47} & {65.09 $\pm$ 5.80} & {96.42 $\pm$ 2.77} \\

{5} & {84.09 $\pm$ 2.73} & {63.95 $\pm$ 5.23} & {94.03 $\pm$ 3.13} \\
\hline
\multicolumn{4}{l}{*With clinical information.}
\end{tabular}
\caption{Model's performance for the extended dataset to compare the impact of clinical data in the classification process}
\label{tab_clin_study}
\end{table}

\subsubsection{Visualization}

\begin{figure}[htbp]
\centering
\includegraphics[width=0.35\textwidth]{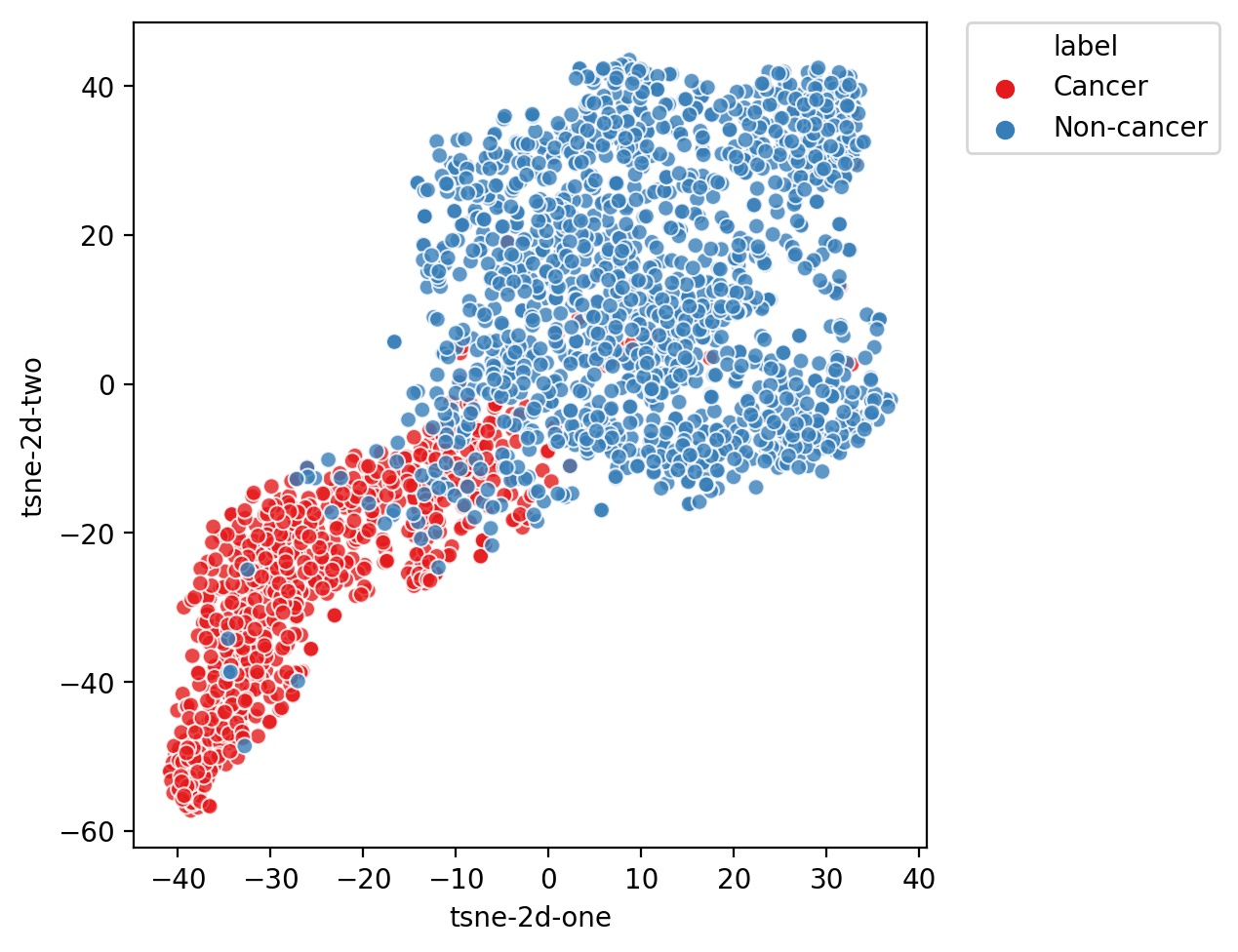}
\caption{Visualization of the features extracted by ResNet50 from all samples using t-SNE.}
\label{fig_tsne}
\end{figure}

We applied the t-Distributed Stochastic Neighbor Embedding (t-SNE) \cite{Maaten2008}, which is a tool to visualize high-dimensional data. In total, 2048 features were extracted from all dataset samples after the last ResNet50 convolutional layer. These features were reduced to two dimensions using t-SNE and shown in Figure \ref{fig_tsne}. From the figure, we can observe that some samples of cancer are overlapped with non-cancer ones.

\begin{figure}[!ht]
\centering
\begin{subfigure}{.24\textwidth}
  \centering
  \includegraphics[width=.98\linewidth]{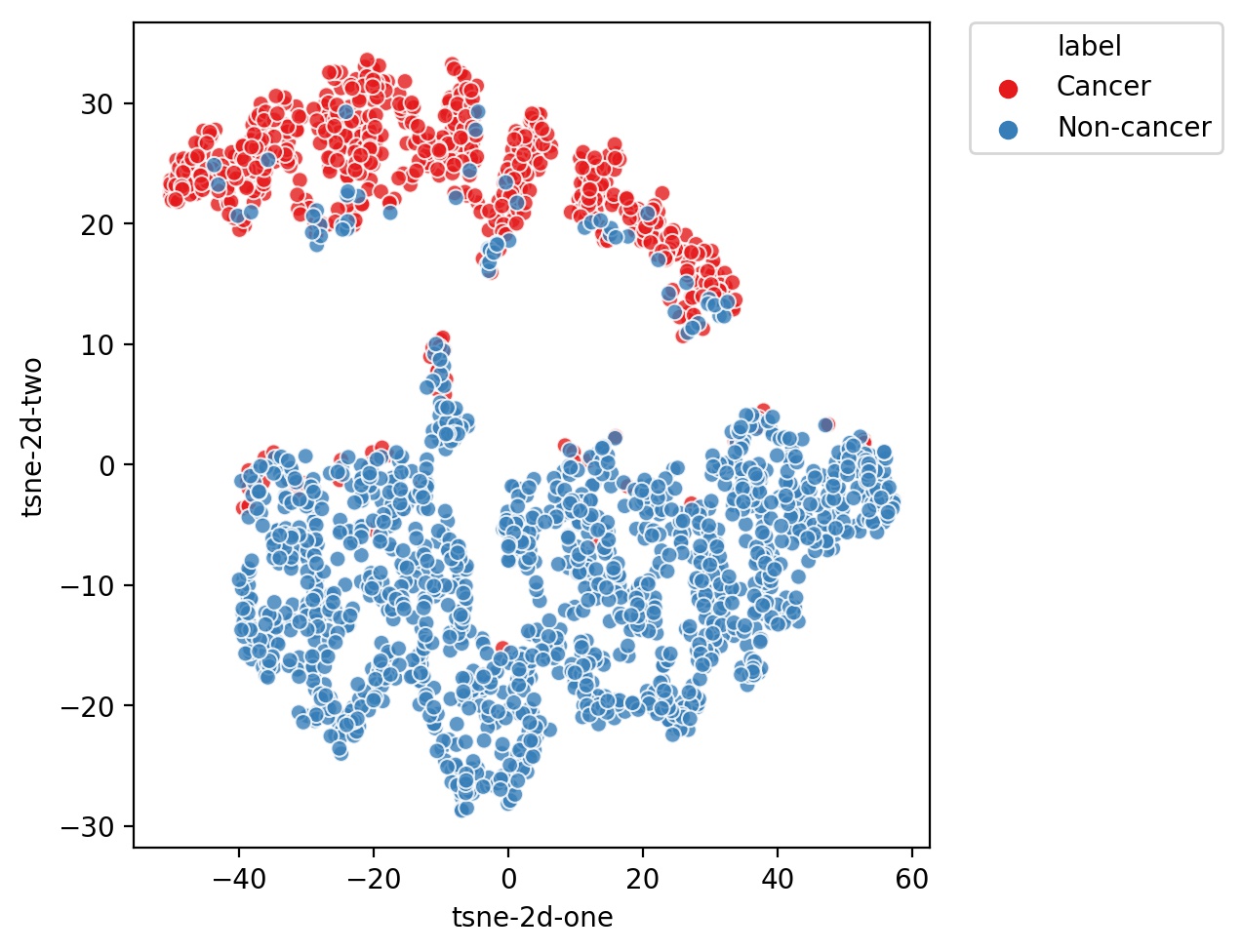}
  \caption{With clinical information}
  \label{fig_tsne_cdc}
\end{subfigure}%
\begin{subfigure}{.24\textwidth}
  \centering
  \includegraphics[width=.98\linewidth]{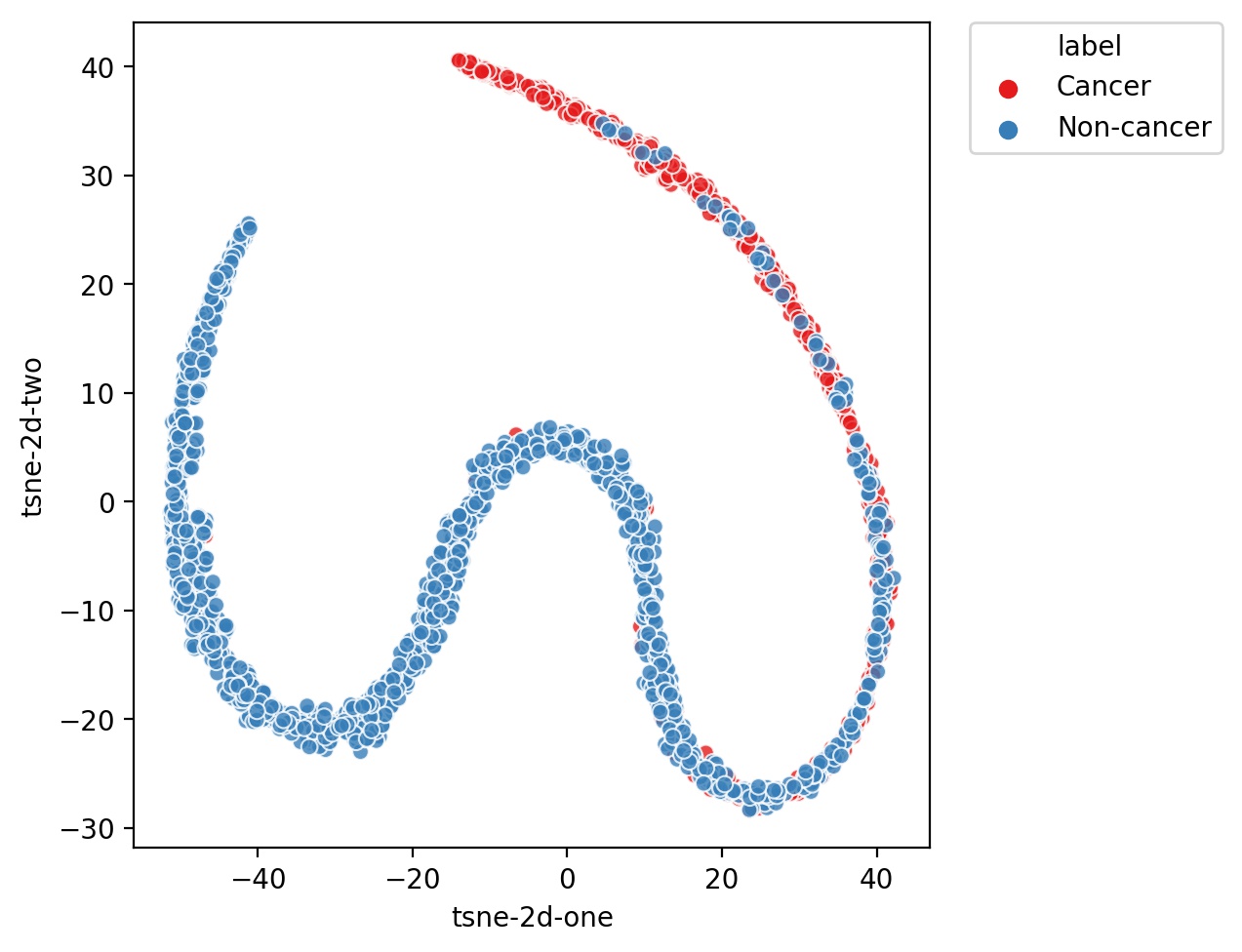}
  \caption{Without clinical information}
  \label{fig_tsne_sdc}
\end{subfigure}
\caption{Visualization of the features extracted by ResNet50 from all samples using t-SNE for both networks}
\label{fig_tsne_clinicos}
\end{figure}

For each network, all the samples were used in order to generate a matrix containing all the reduced features by ResNet50 before the classification layer. From the reduced features matrix, a visualization of these features was obtained using t-SNE and shown in Figure \ref{fig_tsne_clinicos}. From Figure \ref{fig_tsne_clinicos}, it is possible to notice the difference in the representation of the features. 

\subsubsection{The impact of data balancing}

The impact of balancing techniques is assessed by comparing the 2 balancing approaches, i.e., weighted loss function (WGT) and DE. The setup used was the best found in Sec. \ref{subsec_best_setup}. Table \ref{tab_bal} presents the obtained results. The results with the weighted loss function presented the best result in terms of balanced accuracy and recall. The approach based on the mutation operator of DE provided the best result in terms of precision.

\begin{table}[ht!]
\begin{center}
\begin{tabular}{cccc}
\hline
\multirow[c]{2}{*}{\textbf{BAL}} & \multicolumn{3}{c}{\textbf{Metrics}} \\

& \textbf{\textit{BACC}} & \textbf{\textit{PR}}& \textbf{\textit{REC}} \\
\hline
{WGT} & {85.50 $\pm$ 2.47} & {65.09 $\pm$ 5.80} & {96.42 $\pm$ 2.77} \\

{DE} & {84.84 $\pm$ 5.63} & {65.27 $\pm$ 9.65} & {96.12 $\pm$ 2.60} \\
\hline
\end{tabular}
\caption{Model's performance for the extended dataset for each data balance method}
\label{tab_bal}
\end{center}
\end{table}

\subsection{Discussion}
Observing the results in the previous section, we presented an app that may help clinicians with no dermatological experience. The experiment regarding cancer and non-cancer classification, indicates that it can be a promising tool in the screening process, since our preliminary results achieved a balanced accuracy and recall of $85\%$ and $96\%$, on average. However, those experiments were performed on development process, so further experiments in deployment phase are also necessary.

Regarding the classification problem, the presented results confirms the hypothesis raised by Brinker \textit{et. al} \cite{brinker2018} that patient clinical information tends to improve deep learning performance for skin cancer classification. For our particular case, the use of clinical information improved for all the used metrics, achieving a balanced accuracy of $85.5\%$ and a recall of $96.42\%$ in the best scenario. These results can be corroborated from the difference in the representation of the reduced features using t-SNE with and without clinical information, where the network using clinical information was able to provide more distinguished features than the one without, as shown in Figure \ref{fig_tsne_clinicos}. Nevertheless, we observe that the model achieved a slightly lower performance when compared with Pacheco and Krohling \cite{Pacheco2020} results. It occurs because we grouped together lesions that were confused with each other, without the use of clinical information. For example, the model tends to mistake SEK and NEV, which have similar image features. But when we analysis their clinical information, we notice that NEV median age is slower that SEK \cite{Pacheco2020}, which can help in the classification task. Since they were grouped under the same label, non-cancer, the addition of the clinical information does not represent a real gain on the classification as reported in Pacheco and Krohling \cite{Pacheco2020}.

Regarding the impact of data balancing, the approach based on the mutation operator of DE provided similar results to those obtained using weighed loss function.

\section{Conclusion}
In this paper, we presented a smartphone based application to support the diagnostic of skin cancer  using convolutional neural networks. The results obtained with clinical information presents an average balanced accuracy of $85\%$ and a recall of $96\%$. The study of the impact of clinical information has shown that clinical information is relevant to cancer detection since it improved on average the balanced accuracy, precision and recall in about $1.4\%$, $1.1\%$ and $2.4\%$, respectively. Regarding the data balancing approach, the weighted loss function presented the best results but the approach based on the mutation operator of differential evolution is competitive. It is worth mentioning that these results are promising but yet preliminary since our collected dataset is small. The next phase consists of applying this approach to a real world scenario to assist doctors in the screening process. We also continue to collect more data to improve our results.



\section{Acknowledgment}
B. Krohling, P. B. C. Castro, and R. A. Krohling would like to thank the financial support of the Fundação de Amparo a Pesquisa e Inovação do Espírito Santo (FAPES) - grant n. 575/2018. R. A. Krohling also thanks the Conselho Nacional de Desenvolvimento Científico e Tecnológico (CNPq) - grant n.309729/2018-1. We also thank all the members of the Dermatological Assistance Program (PAD-UFES), specially prof. P. L. Frasson, and the support of J. G. M. Esgário, a former Labcin member.

\begin{table}[htbp]

\begin{center}
\begin{tabular}{cccc}
\hline
\multirow[c]{2}{*}{\textbf{Beta}} & \multicolumn{3}{c}{\textbf{Metrics}} \\

& \textbf{\textit{BACC}} & \textbf{\textit{PR}}& \textbf{\textit{REC}} \\
\hline
{1} & {88.12 $\pm$ 1.39} & {81.64 $\pm$ 4.06} & {85.67 $\pm$ 4.07} \\

{3} & {88.23 $\pm$ 3.01} & {75.80 $\pm$ 6.75} & {90.74 $\pm$ 2.57} \\

{5} & {84.09 $\pm$ 2.73} & {63.95 $\pm$ 5.23} & {94.03 $\pm$ 3.13} \\

{7} & {85.83 $\pm$ 1.62} & {68.53 $\pm$ 2.60} & {91.94 $\pm$ 2.42} \\

{10} & {85.83 $\pm$ 3.26} & {68.5 $\pm$ 3.56} & {91.94 $\pm$ 5.05} \\
\hline
\end{tabular}
\caption{Model's performance for F-measure regarding varying beta without clinical data in the classification process}
\label{tab_beta_sem_dados}
\end{center}
\end{table}

\begin{table}[htbp]
\centering

\begin{tabular}{cccc}
\hline
\multirow[c]{2}{*}{\textbf{Beta}} & \multicolumn{3}{c}{\textbf{Metrics}} \\

& \textbf{\textit{BACC}} & \textbf{\textit{PR}}& \textbf{\textit{REC}} \\
\hline
{3*} & {85.50 $\pm$ 2.47} & {65.09 $\pm$ 5.80} & {96.42 $\pm$ 2.77} \\

{5} & {84.09 $\pm$ 2.73} & {63.95 $\pm$ 5.23} & {94.03 $\pm$ 3.13} \\
\hline
\multicolumn{4}{l}{*With clinical information.}
\end{tabular}
\caption{Model's performance for the extended dataset to compare the impact of clinical data in the classification process}
\label{tab_clin_study}
\end{table}

\begin{table}[ht!]
\begin{center}
\begin{tabular}{cccc}
\hline
\multirow[c]{2}{*}{\textbf{BAL}} & \multicolumn{3}{c}{\textbf{Metrics}} \\

& \textbf{\textit{BACC}} & \textbf{\textit{PR}}& \textbf{\textit{REC}} \\
\hline
{WGT} & {85.50 $\pm$ 2.47} & {65.09 $\pm$ 5.80} & {96.42 $\pm$ 2.77} \\

{DE} & {84.84 $\pm$ 5.63} & {65.27 $\pm$ 9.65} & {96.12 $\pm$ 2.60} \\
\hline
\end{tabular}
\caption{Model's performance for the extended dataset for each data balance method}
\label{tab_bal}
\end{center}
\end{table}

\newpage

\bibliographystyle{IEEEtran}
\bibliography{references,IEEEabrv}





\end{document}